\documentclass{aastex}

\usepackage{emulateapj5}

\shorttitle{Radio Lobe Filled Cosmic Web at High-z}
\shortauthors{Gopal-Krishna \& Wiita}

\begin{document}
\title{Was the Cosmic Web of Protogalactic Material Permeated by
Lobes of Radio Galaxies During the Quasar Era?} 

\author{Gopal-Krishna}
\affil{National Centre for Radio Astrophysics, 
TIFR, Pune University Campus,  Pune 411007, 
India}
\email{krishna@ncra.tifr.res.in}
\and
\author{Paul J.\ Wiita\altaffilmark{1}}
\affil{Department of Physics \& Astronomy, Georgia State University,
Atlanta, GA 30303-3083}
\email{wiita@chara.gsu.edu}

\altaffiltext{1}{also, Department of Astrophysical Sciences, Princeton
University}

\begin{abstract}

Evidence for extended active lifetimes ($> 10^8$ yr) for radio galaxies
 implies that many large radio lobes were produced during the 
``quasar era'', 
$1.5 < z < 3$, when the comoving density of radio sources was
2 -- 3 dex higher than the present level.  However, 
inverse Compton losses against the intense microwave
background
substantially reduce the ages and numbers of sources that are detected in flux-limited 
surveys. 
The realization that the galaxy forming material in those epochs 
was concentrated in filaments occupying
a small fraction of the total volume then leads to the conclusion that 
radio lobes permeated much of the volume occupied by the protogalactic material during 
 that era.  The sustained overpressure in these extended lobes 
is likely to have played an important role in triggering the high 
inferred rate of galaxy formation at $z > 1.5$ and in the magnetization of the 
cosmic network of filaments.

\end{abstract}


\section{Introduction}

Studies of flux density limited samples of radio sources have revealed
that the comoving space densities of powerful radio galaxies declined
between two and three orders of magnitude from redshifts of 2--3
to the present epoch (Dunlop \& Peacock 1990;
Jackson \& Wall 1999; Jarvis \& Rawlings 2000; 
Rawlings 2001; Willott et al.\ 2001).  
The star formation rate as deduced from near-infrared and optical surveys 
rises with cosmic epoch in a still poorly determined
fashion (e.g. Steidel et al.\ 1999) until $z \sim 2$, and 
subsequently begins to decline at $z < 1-2$ (Connolly et al.\ 1997; 
Madau, Pozzetti \& Dickinson  1998;  Cowie, Songaila \& Barger 1999). However, recent studies indicate that the peak
of comoving sub-mm derived luminosity density occurred at $z \sim 2-5$
and was considerably greater than that inferred from optical surveys,
which are probably affected by dust obscuration
(e.g.\ Blain et al.\ 1999; Archibald et al.\ 2001).  In view of
these similar trends, it is of great interest to examine any
possible links between these evolutionary patterns. 
Several authors  have studied the role of active galaxies
in compressing  intergalactic clouds by their expanding
radio jets or lobes, which could trigger large-scale star formation in
the circumgalactic material (e.g.\ Begelman \& Cioffi 1989; De Young 1989; 
Rees 1989; Chokshi 1997).
A common perception is that despite the strong cosmological
evolution of the number density of radio galaxies they fill only a  
 minute fraction of the volume of the high-$z$
universe.  In this {\it Letter} we point out that the volume filling factor
of the lobes of radio galaxies may well have been grossly underestimated.

The role of inverse Compton (IC) losses due to the cosmic microwave background
radiation (CMBR) 
in limiting the observed life spans of high-$z$ radio galaxies has been emphasized
recently by Blundell \& Rawlings (1999) and Blundell, Rawlings \& Willott
(1999, hereafter BRW99) although it has been considered in earlier studies
(e.g.\ Rees \& Setti 1968; Gopal-Krishna, Wiita \& Saripalli 1989;  
Kaiser, Dennett-Thorpe \& Alexander 1997).  
Thanks to
both adiabatic and IC losses, a significant fraction of sources, both intrinsically weak ones,
 as well
as the older, and hence,
most expanded, intrinsically powerful ones,  would be rendered undetectable
in flux limited surveys for much of their active lifetimes (\S 2.1).   
In addition, the redshift interval over which
radio galaxies were much more abundant, centered around $z \simeq 2.5$,
the so-called ``quasar era'', lasted long enough
to encompass several generations of
radio active galaxies
(e.g.\ Jarvis \& Rawlings 2000; Barger et al.\ 2001) (\S 2.2).

The intense galaxy formation activity
during that era was confined to the sheets and filaments
containing higher density baryonic material, much in the form of the
warm-hot medium (with temperatures of $10^{5-7} {\rm K}$).  As
indicated by recent simulations (Cen \& Ostriker 1999) such 
filamentary structures probably occupied
only a tiny fraction ($\eta \sim 0.03$) of the universe at
$z \simeq 2.5$. By that time, most galaxies already formed, along with
much circumgalactic material, would be located at the junctions
of the filaments
(Bryan \& Norman 1998; Dav{\'e} et al.\  2001) (\S 2.3).
 
Although none of the above
factors are precisely known,  we argue that
reasonable estimates for all of them can be obtained, and thus infer
that much  of
the filamentary structure containing the protogalactic medium was 
directly impacted by the expanding overpressured lobes of 
radio galaxies born during 
the quasar era (\S 3).  Their role in triggering the dramatic star formation activity 
seen at $z \sim~$1--2 should therefore be closely examined.

\section{The Volume Encompassed by Radio Lobes}

\subsection{The truncated visibility of high-$z$ sources}

As highlighted by BRW99, the depletion 
of energetic
particles due to adiabatic and enhanced IC losses 
at high-$z$ can lead to a ``youth-redshift degeneracy'' (Blundell \& Rawlings
1999).  With increasing $z$, radio galaxies will only be observed
in progressively earlier phases in any flux limited survey.
Taking a fixed lifetime $T$, of $5 \times 10^8$ yr for
the nuclear activity, BRW99 have argued that at $z = 2.5$, a source with
a fairly high beam power
of $2 \times 10^{38}$W  would become undetectable after only a few
Myr, even in the  deepest meter-wavelength (151 MHz)
complete samples with redshift data.
We define the fractional duration of detectability, $f_d \equiv \tau/T$,
with $\tau$ the time for which the source remains above the flux density
limit of the survey. 
It should be noted that even after falling below the detection threshold, 
the 
lobes of such a radio source will continue to grow in size.

While we have drawn upon the most recent and sophisticated
models for beam dynamics, 
which successfully reproduces a large number 
of slices in the observed [$P,D,z,\alpha$] parameter space,
the poor understanding of the high-$z$
circumgalactic medium introduces uncertainties, as does the 
neutral assumption of a redshift independent active lifetime, $T$.
In the BRW99 model, the total length of the source, $D(t)$, is related 
to the power in each beam, $Q_0$:
\begin{equation}
D(t) = 2c_1 a_0 \Bigl(\frac{t^3 Q_0}{a_0^5 \rho_0}\Bigr)^{1/(5-\beta)}.
\end{equation}
The ambient density profile is
$\rho_{\rm ext}(r) = \rho_0 (r/a_0)^{-\beta}$, with reasonable values being
$\rho_0 = 1.67 \times 10^{-23}{\rm kg}~{\rm m}^{-3}$, $a_0 = 10~$kpc,
$\beta = 1.5$, and $c_1 = 1.8$.  In the absence of a consensus on the
cosmological evolution of these parameters, we follow BRW99 in making
the neutral assumption that $a_0, \beta$, $\rho_0$ and $T$ are 
independent of $z$; however, we allow for $1 \le T/10^8{\rm y} \le 5$. 
Here we consider cosmologies with $H_0 = 50~ {\rm km~ s}^{-1} 
{\rm Mpc}^{-1}, \Omega_{\rm M} = 0$ or $1$, and $\Omega_{\Lambda} = 0$.

\subsection{Radio luminosity function (RLF)}

The most recent determination of the cosmological evolution of the
RLF is based on three flux-limited samples derived from surveys at
low radio frequencies (151/178 MHz), the 7CRS,
6CE, and 3CRR, which are complete above the flux density limits
ranging from $0.5 - 12$ Jy at 151 MHz (Willott et al.\ 2001).  
Spectroscopic redshifts
 obtained for 96\% of the total of 357 sources in these
samples
 provide an unprecedented coverage of the
$P-z$ plane, and thereby can resolve substantially the degeneracy
 between luminosity, $P$, and $z$.  The selection at low frequencies
minimizes the orientation bias due to relativistic beaming and hence 
the morphological diversity of sources.

Note that the derived RLFs only reflect the sources that are visible
above the 0.5 Jy limit of the deepest sample, so the computed
durations of the visible phase at 151 MHz in the BRW99 model at
can be directly applied to
these RLFs in a self-consistent manner.    
The very rapid cosmological evolution of the more powerful sources,
of Fanaroff-Riley (1974, FR) Type II, with $P_{151} \ge 10^{25.5} 
{\rm W~Hz}^{-1}{\rm sr}^{-1}$  makes the RLF at 
$z \simeq 2.5$ essentially flat between 
that power and log $P_{151} = 27$, above which the RLF declines rapidly.
By $z \sim 2$, the RLF of powerful sources has risen by nearly 3 dex above
the local RLF, followed by a possible slow decline at higher redshifts
(Willott et al.\ 2001; also Jarvis \& Rawlings 2000).

As our main concern is with the volume encompassed by radio 
lobes during the entire quasar era, we shall consider all powerful, i.e., all 
FR II, radio sources, whether  radio galaxies or quasars.
The comoving space density of FR II sources 
around $\log P_{151} = 25.5$ is $\rho_{\rm obs} \simeq [6.3,3.1] \times 10^{-7} ~{\rm Mpc}^{-3}
(\Delta {\log} P_{151})^{-1}$, where, throughout this paper, pairs of
values in square brackets are for
$\Omega_M$ = $1$ and $0$, respectively (Fig.\ 3 of Willott et al.\ 2001).
In order to correct these observed values of $\rho_{\rm obs}$ for the sources
which have fallen below the detection limit, they should be divided
by the appropriate value of $f_d$.  The simulation results
presented in BRW99 (Figs.\ 13 and 14) permit a conservative 
useful estimate of the mean of $f_d$ at $z = 2$.  A more
detailed calculation is underway (Kulkarni, Gopal-Krishna \& Wiita 2001).
We infer that for a source
to appear at all in the BRW99 dataset it must have a 
$Q_0 > 7.5 \times 10^{37}{\rm W}
\equiv Q_m$,
where it will have ${\log} P_{151} = 27.0$ at an early evolutionary stage
 ($\tau \simeq 1~$Myr).
  A  source with $Q_0 = 2 \times 10^{38}$W has 
$\tau \simeq 9$ Myr,
 while one with $Q_0 = 1.3 \times 10^{40}$W has
$\tau \simeq 70$ Myr, so that, roughly, $\tau \propto Q_0^{0.5}$, for
$Q_0 > Q_m$.
Assuming a fixed lifetime for all radio galaxies, $T_5 = (T/500~{\rm Myr})$
(cf. BRW99), a
source will continue to expand its lobes for an extra factor of 
$T/\tau = f_d^{-1}$ times beyond the age at which it falls 
below the detection threshold of their deepest 
sample.  Using the same probability distribution of beam 
powers as empirically inferred  by BRW99, 
$p(Q) dQ_0 \propto Q_0^{-2.6}dQ_0$ for 
$Q_{\rm min} \equiv 5 \times 10^{37}{\rm W} < Q_0 < 
5 \times 10^{42}{\rm W} \equiv Q_{\rm max}$ and 
$p(Q_0) = 0$ otherwise,
we find that, normalizing to $Q_0^* = 1.3 \times 10^{40}$W, for which
$f_d^* = 0.14~T_5^{-1}$ (Fig.\ 13 of BRW99),
$\langle f_d \rangle = 0.11f_d^*  \simeq 1.5 \times 10^{-2}~T_5^{-1}.$ 


This correction factor should be applied to the observed source densities at 
and above the luminosity where the RLF($z = 2$) steepens, that is for 
$\log P_{151} \ge 26.5$ for $\Omega_{\rm M} = 1$ and 
$\log P_{151} \ge 27.0$ for $\Omega_{\rm M} = 0$
(see Fig.\ 3 of Willott et al.\ 2001).
For the remaining FR II sources, i.e., those appearing at lower
luminosities in the RLF at $z = 2$, $f_d$ is clearly
expected to be still smaller, in that such sources will have 
$\tau < 1$ Myr (BRW99).  Furthermore, these estimates of $f_d$ for 
$z=2$ are conservative upper limits for $z = 2.5$, close to the 
 peak of the quasar era (e.g., Rawlings 2001), and where the 
CMBR is even stronger.

In addition, there will be a larger population of radio sources at $z = 2.5$
which are simply not detected, because their powers are too low even 
at early stages, but which
will also cumulatively inflate a substantial lobe volume.  It is known
that starburst activity can occur along the edges of even FR I radio sources
(e.g., McNamara \& O'Connell 1993).
Maintaining our conservative approach, we shall ignore this 
additional contribution.
Now, dividing $\rho_{\rm obs}(1+z)^3$ by the ${\langle f_d \rangle}$ 
gives the actual proper density
at $z = 2.5$ of powerful radio sources born in an interval
$T$, 
\begin{equation}
\rho \simeq [4.1, 2.1] \times 10^{-5}
(1+z)^3 T_5~{\rm Mpc}^{-3} (\Delta \log P_{151})^{-1}
\end{equation}
(Willott et al.\ 2001). 
To obtain the integrated density of radio sources we 
consider the width of the relevant $ {\log} P_{151}$ bin, which is about
[1.25, 1.5] dex. 
Thus the total proper
density of galaxies with beam powers sufficient to produce FR II sources 
(whether or not they are detected in the survey) is $\phi(T) = [5.1,3.1]  \times 10^{-5}
T_5 (1+z)^3 {\rm Mpc}^{-3}$. 

We must finally account for the fact that the epoch during which the  number
density of sources is roughly constant at the above value extends from 
$z \simeq 1.5$ to $z \simeq 3$, with characteristic $z \simeq 2.5$ 
(Jarvis \& Rawlings 2000; Rawlings 2001).  
This corresponds to a quasar era of length $t_{\rm QE} \sim 2$ Gyr 
which encompasses several generations
of radio sources.
The values of $t_{\rm QE}$ vary with $\Omega_M$ so as to compensate 
for the difference due to cosmology in the definition
of $\rho_{\rm obs}$, so we finally find that
the total proper density, $\Phi$, of intrinsically powerful 
radio sources is essentially independent of $T$ (as long as it 
exceeds $\sim 10^8$ yr) and $\Omega_M$:
$\Phi = \phi(T) (t_{\rm QE}/ T) = 7.7 \times 10^{-3} ~{\rm Mpc}^{-3}$.

\subsection{Relevant volume filling factor of radio lobes}

Recent high-resolution hydrodynamic simulations
of $\Lambda$CDM models suggest that at the present epoch roughly 70\% of
the baryons exist in a web of filaments as warm-hot gas and embedded
galaxies and clusters, altogether occupying about 10\% of the
volume of the universe (Cen \& Ostriker 1999; Dav{\'e} et al.\ 2001).  However, at $z \simeq
2.5$, the network of filaments occupied only around 3\% 
of the comoving volume, and
their mass content has steadily grown since that epoch from about 20\%,
at the expense of the surrounding warm medium (the  
gas cooler than $\sim~10^5$K, responsible
for the Lyman-${\alpha}$ absorption).

Since massive galaxies, the progenitors of powerful radio sources, lie
near the junctions of the filaments, their radio jets and lobes are expected 
to directly
interact with the cool circumgalactic material as well as the warm-hot
and hot gas contained in the filaments.  Significant amounts of star 
formation are triggered by the shocks and 
high pressure associated with the radio emitting features (\S 3).    
Thus, if a good fraction of this {\it relevant} volume of the universe was 
permeated by
radio lobes in the quasar era,  the lobes could play a substantial 
role in triggering the intense star formation activity seen in the universe
at
$z \sim 1-2$.
 
We can now examine the viability of this proposal.
The effective volume of relevance here is just that of the filaments
containing the galaxies and overdense protogalactic gas at 
$z \sim 2.5$, which is only 
the fraction 
$\eta$ of the total volume.  The volume occupied by the synchrotron emitting
lobes of a powerful radio source (actually a lower limit to  the 
volume encompassed by the outer bow shock) at an age $t$ is 
\begin{equation}
V(t,Q_0)  \simeq (\pi/4) D(t, Q_0)^3 R_T^{-2},
\end{equation}
where $R_T$, the ratio of the source's length, $D$, to its
width, $2R$, is typically
$\sim 5$ (e.g.\ Leahy \& Williams 1984).  
By integrating equation (3), weighted by the
distribution function $p(Q_0)$ (see \S 2.2), and using equation (1) for $D(t, Q_0)$,
we compute the average volume filled by these sources at their maximum ages
to be
$\langle V(T) \rangle = 2.1~ T_5^{18/7} {\rm Mpc}^3$.

\section{Discussion and Conclusions}

Putting together the results from \S 2, 
the fractional relevant volume which radio lobes born during
the quasar era cumulatively cover is,
\begin{equation}
\zeta = \Phi ~\langle V (5 \times 10^8 {\rm yr}) 
\rangle ~(0.03/\eta)(5/R_T)^2 \simeq 0.5,
\end{equation}
for our canonical choice of $T$ (BRW99).  
We emphasize that this filling factor is the sum of the lobe volumes created during the entire quasar era; this is relevant for estimating the
domain of star formation triggered by the lobes.  In contrast,
only one generation of sources is considered in estimating the
contribution to the  energy density, $u$, of synchrotron plasma injected into
the cosmic web by their lobes, since the left-over contribution from
previous generations of lobes
should be marginal,  due to severe expansion losses.
This leads to $u \simeq 2.7~ T Q_m \phi(T) \approx 2 \times 10^{-16} {\rm J~m}^{-3}$ within the filaments. 
See Table 1 for specific values. 


Thus, the main result is that, quite plausibly, a very significant 
fraction of the relevant
volume of the universe was impinged upon by the growing
radio lobes during the redshift interval when radio source production was at
its peak ($z \simeq 2.5$).   
Radio lobes propagating through
this protogalactic medium mainly encounter the hot ($T > 10^6$K), 
volume filling, lower density 
gas,  but when they envelop the 
embedded cooler clumps ($T \sim 10^4\,$K; Fall \& Rees 1985), 
the initial bow shock compression
will trigger large-scale star formation, which is sustained by the persistent
overpressure from the engulfing radio cocoon.  Note that the cocoon pressure
is likely to be well above the equipartition estimate (Blundell \&
Rawlings 2000).   
This scenario is supported by many models,
both analytical (e.g.\ Begelman \& Cioffi 1989; Rees 1989; Daly 1990), and
hydrodynamical (e.g.\ De Young 1989; Cioffi \& Blondin 1992), 
and provides an explanation for the remarkable radio--optical 
alignment effect exhibited by high-$z$
radio galaxies (e.g., McCarthy
et al.\ 1987; Chambers, Miley \& van Breugel 1988).  Additional support for
jet or lobe-induced star formation comes from the HST images 
of $z\sim 1$ radio galaxies (Best, Longair \& R{\"o}ttgering 1996),
and of some radio sources at higher $z$ (e.g.\ Miley et al.\ 1992;
Bicknell et al.\ 2000).

It is important to check if the overpressure of the expanding
lobes over the ambient medium persists throughout the active lifetime of
the radio source.  From BRW99 (cf.\ Falle 1991):  
$p_{\rm lobe} \propto  
t^{(-4-\beta)/(5-\beta)}$, but $D \propto t^{3/(5-\beta)}$, so
$p_{\rm lobe} \propto D^{(-4-\beta)/3}$.  The external pressure
declines less rapidly, $p_{\rm ext} \propto D^{-\beta}$, so
$p_{\rm lobe}/p_{\rm ext} \propto D^{(-4+2\beta)/3}$.  For
$\beta = 3/2$,  $p_{\rm lobe}/p_{\rm ext} \propto D^{-1/3}$, 
while for $\beta = 1$, which might be more reasonable at large radial
distances,
$p_{\rm lobe}/p_{\rm ext} \propto
D^{-2/3}$.  For the ranges of $Q_0$, $\rho_0$ and $a_0$ considered here,
appropriate for FR II sources, overpressures at $D = 50$ kpc will amount 
to factors of $10^2$--$10^4$, corresponding to Mach numbers of 10--100
(BRW99) for the bowshock.  Thus, overpressure should persist 
even for $D \gg 1$ Mpc, sustaining lobe expansion even after the jet activity ceases. 
Supersonic expansion into a two-phase circumgalactic medium will compress 
many of the cooler clouds, rapidly reducing
the Jeans mass by factors of 10--100 and thereby 
triggering starbursts (Rees 1989).
Even if most of the gas is in a single
phase, rapid cooling behind the bow shock can 
trigger star formation (Daly 1990).  The lobe-induced star formation
would clearly be concentrated around the site of the active galaxy,
thereby naturally introducing bias in the distribution of star forming sites.

Chokshi (1997) 
has suggested that radio sources at high redshift ($2 < z < 3$),
powered by accretion of protogalactic material onto preexisting  
black holes, assembled their host elliptical galaxies via radio lobe induced 
starbursts.  She argued for such an origin of the entire population 
of massive ellipticals seen at the present epoch.
Here  our focus is on quantitatively assessing if radio sources 
could have made an impact on the entire star/galaxy formation process 
on the global scale.
Keeping a conservative bias, we have only considered FR II lobes and
therefore used a value for $\Phi(z=2.5)$ 
which is less than a tenth of the value used by
Chokshi.   Inspired by recent models 
of radio source evolution and cosmic structure formation, we have 
taken into account a number of additional
factors, which yield significant levels of effective volume coverage 
attained by radio galaxy lobes during the cosmic era of peak radio 
source production.

While we have focused our attention on the ``quasar era'' it is clear
that $f_d$ for radio galaxies at $z > 3$ would be yet
smaller; hence $\rho$ and their fractional volume
coverage may not decline for $z > 3$, even
if the quasar population was sparser,
as deduced by Shaver et al.\ (1998).  But the data 
are inadequate to rule out very slow declines at $z > 3$
 (e.g., Jarvis \& Rawlings 2000).  Thus, even at those
very early epochs, radio sources may have contributed substantially to the
formation of galaxies,  which have recently been
discovered from IR and sub-mm surveys 
(e.g., Steidel et al.\ 1999; Blain et al.\ 1999; Archibald et al.\ 2001).

An interesting corollary of our picture is that the radio lobes 
could efficiently seed with magnetic field
a large portion of the cosmic web (cf.\ Daly \& Loeb 1990).  From $u$ (Table 1)
we estimate an equipartition magnetic field in the filaments 
to be $\sim10^{-8}$ G.  This relatively strong field is in accord with the 
recent estimate for the cosmic web of filaments based on a more realistic 
interpretation of rotation measure data (Ryu, Kang \& Biermann 1998).

We have ignored the relatively small overpressured radio lobes
associated with the weaker, albeit more abundant, FR I sources. 
We have also neglected the explicit growth of
three dimensional instabilities which will afflict even very
powerful jets trying to propagate to Mpc distances (e.g., Hooda \& Wiita 1998); however, under these circumstances the jet advance is
likely to resemble the ``dentist drill'' scenario of Scheuer (1982),
so while the hotspot emission may weaken, the lobes can continue 
to inflate and expand.
Both of these effects would tend to further reduce  the value 
of $f_d$ and thereby increase our estimates for $\zeta$.  However, 
recurrent periods of activity in individual 
galaxies, even if they add up
to $\sim 5 \times 10^8$ yr, should inflate smaller total volumes.  These
details will be explored in future work.

\acknowledgments
We thank Vasant Kulkarni, Arun Mangalam and Rajaram Nityananda for helpful comments.  
PJW is grateful for support from Research Program Enhancement funds at GSU.


\begin{deluxetable}{c c c c c c c}
\tabletypesize{\footnotesize}
\tablewidth{8.5 cm}
\tablecaption{Densities, volumes and filling factors}
\tablehead{
\colhead{$T$\tablenotemark{a}} & \colhead{$\Omega_M$} & \colhead{$\langle f_d\rangle$} &
\colhead{$\phi(T)$\tablenotemark{a}} & \colhead{$\langle V(T) \rangle$\tablenotemark{a}} & 
\colhead{$\zeta$} & \colhead{$u$\tablenotemark{a}}
}
\startdata
100 & 0 & 0.077  & 6.2($-$6)  & 0.033 & 0.01 & 5.6($-$18)\\
100 & 1 & 0.077  & 1.0($-$5)  & 0.033 & 0.01 & 9.2($-$18) \\
300 & 0 & 0.026  & 1.9($-$5)  & 0.55  & 0.08 & 5.0($-$17)\\
300 & 1 & 0.026  & 3.1($-$5)  & 0.55  & 0.08 & 8.2($-$17)\\
500 & 0 & 0.015  & 3.1($-$5)  & 2.05  & 0.53 & 1.4($-$16)\\
500 & 1 & 0.015  & 5.1($-$5)  & 2.05  & 0.53 & 2.3($-$16)\\

\enddata
\tablenotetext{a}{Units: $T ({\rm Myr)}; \phi(T)({\rm Mpc}^{-3}); \langle V(T) \rangle({\rm Mpc}^3); u({\rm J m}^{-3})$}
\end{deluxetable}


\begin{references}

\reference {} Archibald, E.\ N., Dunlop, J.\ S., Hughes, D.\ H., Rawlings, S., Eales, 
S.\ A., \& Ivison, R.\ J. 2001, MNRAS, 323, 417

\reference {} Barger, A.\ J., et al. 2001, submitted to AJ (astro-ph/0106219)

\reference {} Begelman, M.\ C., \& Cioffi, D.\ F. 1989, ApJ, 345, L21

\reference {} Best, P.\ N.,  Longair, M.\ S., \& R{\"o}ttgering, H.\ J.\ A. 1996, MNRAS, 280, L9

\reference {} Bicknell, G.\ V., Sutherland, R.\ S., van Breugel, W.\ J.\ M.,
Dopita, M.\ A., Dey, A., \& Miley, G.\ K. 2000, ApJ, 540, 678

\reference {} Blain, A.\ W., Smail, I., Ivison, R.\ J., \& Kneib, J.-P. 1999, 
MNRAS, 302, 632

\reference {} Blundell, K.\ M., Rawlings, S., \& Willott, C.\ J. 1999, AJ, 677 (BRW99)

\reference {} Blundell, K.\ M., \& Rawlings, S. 1999, Nature, 399, 330

\reference {} Blundell, K.\ M., \& Rawlings, S. 2000, AJ, 119, 1111

\reference {} Bryan, G.\ L., \& Norman, M.\ L. 1998, ApJ, 495, 80


\reference {} Cen, R., \& Ostriker, J.\ P. 1999, ApJ, 519, L109

\reference {} Chambers, K.\ C., Miley, G.\ K., \& van Breugel, W. 1988, ApJ,
327, L47

\reference {} Chokshi, A. 1997, ApJ, 491, 78

\reference {} Cioffi, D.\ F., \& Blondin, J.\ M. 1992, ApJ, 392, 458

\reference {} Connolly, A.\ J., Szalay, A.\ S., Dickinson, M., SubbaRao, M.\ U.,
\& Brunner, R.\ J.  1997, ApJ, 486, L11

\reference {} Cowie, L., Songaila, A.,  \& Barger, A.\ J. 1999, AJ, 118, 603

\reference {} Daly, R.\ A. 1990, ApJ, 335, 416

\reference {} Daly, R.\ A., \& Loeb, A. 1990, ApJ, 364, 451


\reference {} Dav{\'e}, R., et al.  2001, ApJ, in press (astro-ph/0007217)

\reference {} De Young, D. S. 1989, ApJ, 342, L59 

\reference {} Dunlop, J.\ S., \& Peacock, J.\ A. 1990, MNRAS, 247, 19

\reference {} Fall, M.\ S., \& Rees, M.\ J. 1985, ApJ, 298, 18

\reference {} Falle, S.\ A.\ E.\ G. 1991, MNRAS, 250, 581

\reference {} Fanaroff, B.\ L., \& Riley, J.\ M. 1974, MNRAS, 167, 31P

\reference {} Gopal-Krishna, Wiita, P.\ J., \& Saripalli, L. 1989, MNRAS, 239, 173 

\reference {} Hooda, J. S., \& Wiita, P. J. 1998, ApJ, 493, 81

\reference {} Jackson, C., \& Wall, J.\ V. 1999, MNRAS, 304, 160

\reference {} Jarvis, M.\ J., \& Rawlings, S. 2000, MNRAS, 319, 121


\reference {} Kaiser, C.\ R., Dennett-Thorpe, J., \& Alexander, P. 1997, MNRAS, 292, 723

\reference {} Kulkarni, V.\ K., Gopal-Krishna, \& Wiita, P.\ J. 2001, in preparation

\reference {} Leahy, J.\ P., \& Williams, A.\ G. 1984, MNRAS, 210, 929

\reference {} Madau, P., Pozzetti, L., \& Dickinson, M.  1998, ApJ, 498, 106

\reference {} McCarthy, P. J., van Breugel, W.\ J.\ M., Spinrad, H., \& 
Djorgovski, S. 1987, ApJ, 321, L29

\reference {} McNamara, B.\ R., \& O'Connell, R.\ W.  1993, AJ, 105, 417

\reference {} Miley, G., Chambers, K., van Breugel, W., \& Macchetto, F. 1992, ApJ, 401, L69

\reference {} Rawlings, S. 2001, in IAU Symp.\ 199, The Universe at Low 
Radio Frequencies,  ed.\ Rao, A.\ P., Swarup, G., \& Gopal-Krishna (San Francisco: 
ASP), in press

\reference {} Rees, M.\ J. 1989, MNRAS, 239, 1P

\reference {} Rees, M.\ J., \& Setti, G. 1968, Nature, 219, 127

\reference {} Ryu, D., Kang, H., \& Biermann, P.\ L. 1998, A\&A, 335, 19

\reference {} Scheuer, P.\ A.\ G. 1982, in Extragalactic Radio Sources, ed.\
D.\ S.\ Heeschen \& C.\ M.\ Wade (Dordrecht: Reidel), 21

\reference {} Shaver, P.\ A., Wall, J.\ V., Kellermann, K.\ I., Jackson, C.\ A.,
\& Hawkins, M.\ R.\ S. 1996, Nature, 384, 439

\reference {} Steidel, C.\ C., Adelberger, K.\ L., Giavalisco, M., Dickinson,
M., \& Pettini, M.  1999, ApJ, 519, 1

\reference {} Willott, C.\ J., Rawlings, S., Blundell, K.\ M., Lacy, M., \& Eales, S.\ A. 2001, MNRAS, 322, 536

\end{references}
\end{document}